# When Experts Disagree: Characterizing Annotator Variability for Vessel Segmentation in DSA Images

M. Geshvadi[a,b], G. So[b,c], D.D. Chlorogiannis[b], C. Galvin[b], E. Torio[b], A. Azimi[b], Y. Tachie-Baffour[b],
N. Haouchine[b,d], A. Golby[b,d], M. Vangel[b], W.M. Wells[b,d], Y. Epelboym[b,d], R. Du[b,d], F. Durupinar[a], S. Frisken[b,d]

[a]University of Massachusetts, Boston, MA, USA; [b]Brigham and Women's Hospital, Boston, MA, USA; [c]University of Waterloo, Waterloo Ontario, Canada; [d]Harvard Medical School, Boston, MA, USA

## 1. DESCRIPTION OF PURPOSE

Digital Subtraction Angiography (DSA) is a widely used imaging modality for assessing intracranial circulation. Accurate vessel segmentation in DSA images is essential for diagnosis, surgical simulation, surgical planning, monitoring interventional procedures, and quantitative evaluation post-surgery. While there have been recent advancements in the segmentation of larger vessels (e.g., see [1-3] and reviews in [4-10]), most existing tools fail to segment small vessels less than 1-2 pixels in diameter [11]. There is a growing need for segmenting thin vessels that is driven by advances in imaging technology and advancements in clinical interventions. Thin vessels play a critical role in identifying tumor-feeding vessels, mapping collateral circulation in stroke patients, planning stent deployment in aneurysm treatment to ensure that small vessels are not inadvertently occluded, and planning for brain–computer interface (BCI) implantation.

Manual segmentation remains the gold standard in clinical and research settings, but it is time-consuming (requiring tens of minutes to hours per image for complex neurovascular structure [11]) and is dependent on the person doing the segmentation (i.e., the *annotator*). Automated segmentation appears to be a promising alternative. However, training reliable AI models often requires <u>ground truth</u> annotations. While it is common in medical image processing to use manual segmentation as ground truth, true ground truth of interior structure cannot be obtained from a living patient because imaging is limited by spatial resolution and is subject to artifacts, and manual segmentation is subject to both inter- and intra-rater variability. To emphasize this important point, we refer to manual segmentation as <u>pseudo ground truth</u> herein.

In this work, we systematically investigate intra- and inter-annotator variability in manual segmentation of blood vessels in 2D DSA images. We characterize the uncertainty in pseudo ground truth segmentations from multiple annotators. We propose methods for using this uncertainty characterization to indicate where additional annotations or expert review may be necessary. In addition, we propose incorporating this uncertainty into loss functions for training robust, uncertainty-aware AI segmentation models [12, 13].

As part of this work, we will release an open-source database of ~2000 image patches from clinical DSA of the brain and corresponding annotations from two expert annotators. We will provide annotations and ratings on a subset of these patches from 5 additional annotators. This work greatly extends existing publicly available annotated DSA images. Finally, we provide open-source software developed for this work to quantify uncertainty in segmentations from multiple annotators.

## 2. METHODS

**Data description and preparation**

Our goal is to develop high quality training data for segmenting small vessels in 2D DSA of the brain. We assembled 66 2D DSA images from 11 patients acquired over a two-year time span at Brigham and Women's Hospital. Images were acquired as part of routine clinical care on a Siemens ARTIS scanner and were accessed retrospectively under IRB approval. This dataset consists of both anterior-posterior (AP) and lateral projections acquired after contrast injection into either the left or right internal carotid artery. For each DSA sequence, a representative image was selected from the early, middle, and late arterial phases. Image size was 960 x 960 pixels for AP projections and 1440 x 1440 for lateral projections. Image resolution was 0.3mm x 0.3mm for AP projections and 0.15mm x 0.15mm for lateral projections. Two expert annotators (A1 and A2) manually contoured the vessels in the 66 images using the VESCL vessel contouring library [11]. MD A1 is resident in interventional radiology. PhD A2 is a research scientist with expertise in medical image segmentation. Figure 1 shows a typical DSA image with contours drawn using VESCL.

Approximately 2000 128x128 image patches were randomly generated from the 66 annotated DSA images and fully anonymized. Each image patch has two associated annotations in three different representations: a binary segmentation, a 2D image of signed distance to vessel edges, and a 2D image of unsigned distance to vessel centerlines.

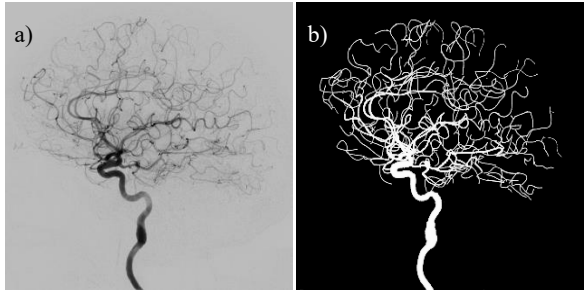

Fig. 1. a) DSA of the right interior carotid artery at mid-arterial phase. b) Vessels of a) segmented using the VESCL contouring library.

As expected for the complex vasculature of the brain, the annotations were not in complete agreement. Initial study showed that we could not simply choose one annotator over the other or combine the two annotations to achieve ground truth. Thus, we performed a set of experiments to investigate regions of disagreement and characterize annotation uncertainty.

**Rating annotations using multiple raters**

We randomly selected 100 patches from our data set of 2000 patches. In each patch, we circled a region and asked 11 raters to respond the question *"Do you think this is a vessel?"*. Raters could respond only with '**Yes**' or '**No**'. The 11 raters included 6 MDs (including 3 neurosurgeons and 2 radiologists) and 5 experts in medical image segmentation. The patches were grouped into four categories:

- NONE (n = 10): Regions where neither annotator identified a vessel
- BOTH (n = 30): Regions where both A1 and A2 annotated the vessel
- A1 Only (n = 30): Regions where only A1 segmented a vessel
- A2 Only (n = 30): Regions where only A2 segmented a vessel

Seven of the patches were randomly duplicated and rotated to measure intra-rater consistency. Annotators A1 and A2 were included in the 11 raters to investigate the consistency of their ratings with their segmentations.

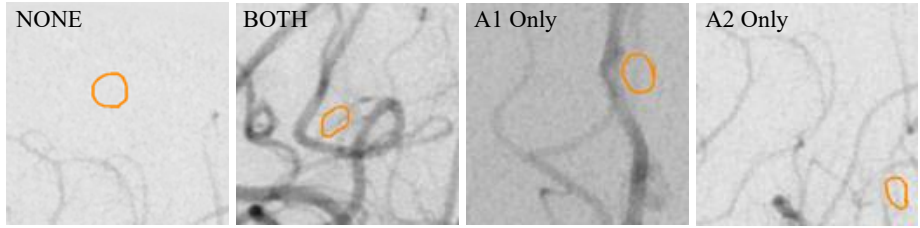

Fig. 2. Four patches representing the four categories.

**Evaluating annotations quantitatively**

*Centerline Dice Coefficient (clDice)*

We used the Centerline Dice Coefficient (clDice) metric described in [14] to quantify agreement between annotators. clDice is a connectivity-preserving similarity metric specifically designed for evaluating tubular and linear structures. It explicitly incorporates each vessel's centerline into the Dice calculation, effectively measuring how well one segmentation's centerline lies within the other's vessel segmentation (and vice versa). clDice is calculated using two values computed from the segmentations: *topological precision* ($T_{prec}$): the proportion of the predicted skeleton that lies within the ground truth mask and *topological sensitivity* ($T_{sens}$): the proportion of the ground truth skeleton that lies within the predicted mask. Then,

$$clDice = \frac{2 \times T_{prec} \times T_{sens}}{T_{prec} + T_{sens}}$$

Conventionally, 1) the skeleton is defined as the set of pixels that lie on the centerline of the structure and 2) the mask is the set of pixels that lie inside the binary segmentation of the structure. When no ground truth is available, clDice is computed between pairs of annotations by assuming that one of the two annotations is ground truth.

For each image or image patch, we extracted centerlines pixels from the 2D image of unsigned distance to vessel centerlines for each segmentation and used the binary vessel segmentation to determine if a pixel on the centerline of one annotation was inside the segmented vessel of the second annotation. We computed clDice for annotators A1 and A2 for each of the 66 annotated DSA images. We also computed clDice pairwise between A1 and A2 and 7 additional annotators for the subset of image patches.

*Modified clDice*

Although clDice was designed to emphasize vessel topology, thereby reducing the inherent bias of the Dice coefficient to favor large structures over small structures, it still favors thicker vessels over thinner vessels. A centerline offset of 1-2 pixels is not problematic for vessels more than 4 pixels in diameter but could result in low clDice scores for vessels less than a few pixels in diameter. To address this bias, we introduce an alternative definition of the mask used for computing $T_{prec}$ and $T_{sens}$. Instead of comparing a first annotator's centerline to a second annotator's *binary segmentation mask*, we redefine the mask as the set of pixels less than a distance threshold from the second annotator's vessel centerlines. We used the 2D image of unsigned distance to vessel centerlines for fast computation of $T_{prec}$ and $T_{sens}$ with this distance-based mask and evaluated sensitivity to the distance threshold. Specifically, we evaluated the modified clDice for distance thresholds ranging from 0.5 to 6 pixels, which correspond to 0.15mm to 1.8mm in AP projections and 0.075mm to 0.9mm in lateral projections.

*Classic filtering methods*

Several classical methods have been used for vessel segmentation. Two of the most common include the Frangi filter [15] and the Sato filter [16]. While these methods have only been partially successful for segmenting complex vasculature and thin vessels, we expect that they could be used to predict image regions where annotators might disagree. To test this hypothesis, we applied multi-scale Frangi and Sato filters available in Scikit-Image [17] and compared the filter responses to 2D images of signed distances to vessel edges for a subset of image patches segmented by 7 annotators.

## 3. RESULTS

**Rating annotations using multiple raters**

Table 1 presents agreement data between raters and annotators on 100 patches. Raters agreed with annotators 90% of the time when A1 and A2 both annotated a vessel and 65% of the time when they both thought a vessel was not present. Raters agreed with A1 76% of the time when A1 thought there was a vessel but A2 did not. Raters agreed with A1 16% of the time when A1 did not think there was a vessel but A2 did. (Reciprocally, raters agreed with A2 24% of the time when A2 thought there was a vessel but A1 did not and raters agreed with A2 84% of the time when A2 did not think there was a vessel but A1 did.)

Table 1. Agreement between raters and annotator A1 when asked if a specified region contained a vessel

| Category (# patches) | Agreement with A1 | | | | | | | | | | Average % agreement |
|---|---|---|---|---|---|---|---|---|---|---|---|
| | R1 | R2 | R3 | R4 | R5 | R6 | R7 | R8 | R9 | R10 | |
| BOTH (30) | 27/30 | 28/30 | 22/30 | 27/30 | 30/30 | 27/30 | 30/30 | 29/30 | 29/30 | 22/30 | 90% |
| A1 Only (30) | 18/30 | 22/30 | 17/30 | 29/30 | 27/30 | 20/30 | 28/30 | 22/30 | 25/30 | 21/30 | 76% |
| A2 Only (30) | 10/30 | 6/30 | 12/30 | 0/30 | 0/30 | 6/30 | 2/30 | 3/30 | 1/30 | 7/30 | 16% |
| NONE (10) | 8/10 | 8/10 | 10/10 | 4/10 | 3/10 | 6/10 | 3/10 | 8/10 | 5/10 | 9/10 | 64% |

**Evaluating annotations quantitatively**

*Centerline Dice Coefficient (clDice)*

Table 2 shows clDice scores comparing annotations by A1 and A2 for the 66 DSA images. Table 3 shows pairwise clDice scores comparing A1 and A2 to 7 additional annotators and for a subset of the 100 128x128 patches.

Table 2. clDice comparing annotations by A1 to A2 for the original 66 DSA images

| | $T_{prec}$ : Mean (Std Dev) | $T_{sens}$ : Mean (Std Dev) | clDice : Mean (Std Dev) |
|---|---|---|---|
| A1 vs A2 | 0.831 (0.065) | 0.843 (0.073) | 0.835 (0.060) |

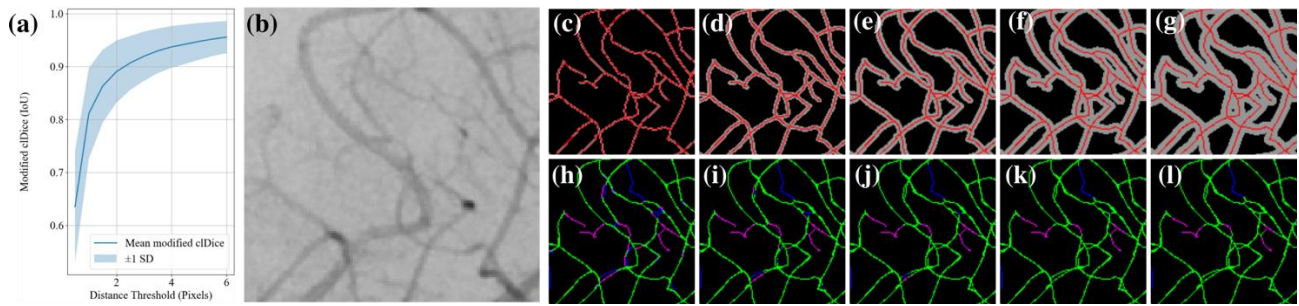

Fig. 3. a) Sensitivity of modified clDice score to the distance threshold for the image patch in 3b). c)-g) centerlines for this patch in red superimposed with the distance mask for a distance threshold of 0.5, 1.5, 2.5, 3.5, and 4.5 pixels. h)-l) shows agreed centerlines (green) centerlines for A1 only (magenta) and A2 only (blue) using distance thresholds of 0.5, 1.5, 2.5, 3.5, and 4.5 pixels respectively.

Table 3. clDice comparing annotations by A1 and A2 to seven additional annotators (a-g) for one image patch

|    | a     | b     | c     | d     | e     | f     | g     | Mean (StdDev) |
|----|-------|-------|-------|-------|-------|-------|-------|---------------|
| A1 | 0.618 | 0.605 | 0.675 | 0.676 | 0.602 | 0.592 | 0.704 | 0.639 (0.041) |
| A2 | 0.759 | 0.739 | 0.803 | 0.821 | 0.689 | 0.734 | 0.822 | 0.767 (0.047) |

*Modified clDice*

Table 4 shows clDice scores computed using the modified distance-based mask for annotators A1 and A2 for a distance threshold of 2.5 pixels. Figure 3 illustrates the sensitivity of the modified clDice scores to the distance threshold. The results change relatively slowly as a function of the distance threshold for distances larger than 2 pixels. Table 5 shows modified (distance-based) clDice scores comparing 7 new annotators to A1 and A2 using a distance threshold of 2.5 pixels.

Table 4. Modified clDice for A1 and A2 annotations in DSA images using a distance threshold of 2.5 pixels

|         | $T_{prec}$ : Mean (Std Dev) | $T_{sens}$ : Mean (Std Dev) | clDice : Mean (Std Dev) |
|---------|-----------------------------|-----------------------------|-------------------------|
| A1 vs A2 | 0.876 (0.055)              | 0.904 (0.051)               | 0.888 (0.038)           |

Table 5. Modified clDice comparing annotations by A1 and A2 to seven additional annotators (a-g) for one image patch

|    | a     | b     | c     | d     | e     | f     | g     | Mean (StdDev) |
|----|-------|-------|-------|-------|-------|-------|-------|---------------|
| A1 | 0.776 | 0.795 | 0.821 | 0.796 | 0.820 | 0.698 | 0.804 | 0.787 (0.042) |
| A2 | 0.888 | 0.907 | 0.904 | 0.882 | 0.903 | 0.792 | 0.895 | 0.881 (0.040) |

## 4. CONCLUSIONS

We investigated inter- and intra-annotator variability in manual segmentations of blood vessels in 2D DSA images. We found significant differences in annotations. These differences contribute to uncertainty in pseudo ground truth segmentations. We presented methods that can be used to quantify this segmentation uncertainty. Finally, we propose ways that segmentation uncertainty can be used to 1) guide additional annotation and 2) develop robust automatic segmentation methods that utilize uncertainty when true ground truth segmentations are not available.

**New or breakthrough work to be presented**

The major novel contributions include 1) a new dataset that consists of ~2000 2D image patches of DSA of blood vessels in the brain together with corresponding manually vessel segmentations from two annotators in 3 formats (binary, distance to vessel centerline, signed distance to vessel edges); 2) an in-depth analysis of uncertainty in the contour data; 3) a process and a set of open-source software tools for evaluating pseudo ground truth in medical image segmentation.

**Statement of originality**

This paper describes original work. It has not been submitted elsewhere.

**Acknowledgements**

This work was funded in part by NIH grants R01EB034223 and R01EB032387.